\documentclass[reprint, prl, superscriptaddress, floatfix, noeprint]{revtex4-1}

\hypersetup{
    colorlinks=true,
    linkcolor=blue,
    citecolor=red,      
    urlcolor=blue,
    pdftitle={Overleaf Example},
    pdfpagemode=FullScreen,
    }
\usepackage{graphicx}
\usepackage{amsmath}
\usepackage{wasysym}
\usepackage{amsfonts}
\usepackage{color}
\usepackage{verbatim}
\usepackage{upgreek}
\usepackage{bm}
\usepackage{textcomp}
\usepackage[symbol*]{footmisc}
\usepackage{float}
\usepackage{soul}
\usepackage{xcolor}

\makeatletter
\newcommand*{\balancecolsandclearpage}{%
  \close@column@grid
  \clearpage
  \twocolumngrid
}
\makeatother

\DefineFNsymbolsTM{otherfnsymbols}{
  \textdagger    \dagger
  \textdaggerdbl \ddagger
  \textsection   \mathsection
  \textparagraph \mathparagraph
 }
\setfnsymbol{otherfnsymbols}

\newcommand{\beq}{\begin{equation}}
\newcommand{\eeq}{\end{equation}}
\newcommand{\bea}{\begin{eqnarray}}
\newcommand{\eea}{\end{eqnarray}}
\newcommand{\bal}{\begin{align}}
\newcommand{\eal}{\end{align}}

\newcommand{\fig}[1]{Fig.\,\ref{#1}}

\newcommand{\WS}{WS$_2$}

\begin{document}
\title{Distance dependence of the energy transfer mechanism in WS\texorpdfstring{$_2$}{2} - graphene heterostructures}
\author{David Tebbe} 
\email{david.tebbe@rwth-aachen.de}
\author{Marc Sch\"utte}  
\affiliation{2nd Institute of Physics and JARA-FIT, RWTH Aachen University, 52074 Aachen, Germany}
\author{K.~Watanabe}
\affiliation{Research Center for Electronic and Optical Materials, National Institute for Materials Science, 1-1 Namiki, Tsukuba 305-0044, Japan}
\author{T.~Taniguchi}
\affiliation{Research Center for Materials Nanoarchitectonics, National Institute for Materials Science, 1-1 Namiki, Tsukuba 305-0044, Japan}
\author{Christoph Stampfer} 
\affiliation{2nd Institute of Physics and JARA-FIT, RWTH Aachen University, 52074 Aachen, Germany}\affiliation{Peter Gr\"unberg Institute (PGI-9), Forschungszentrum J\"ulich, 52425 J\"ulich, Germany}
\author{Bernd Beschoten} 
\affiliation{2nd Institute of Physics and JARA-FIT, RWTH Aachen University, 52074 Aachen, Germany}
\affiliation{JARA-FIT Institute for Quantum Information, Forschungszentrum J\"ulich GmbH and RWTH Aachen University, 52074 Aachen, Germany}
\author{Lutz Waldecker} 
\email{waldecker@physik.rwth-aachen.de}
\affiliation{2nd Institute of Physics and JARA-FIT, RWTH Aachen University, 52074 Aachen, Germany}

\begin{abstract}
We report on the mechanism of energy transfer in van der Waals heterostructures of the two-dimensional semiconductor \WS\ and graphene with varying interlayer distances, achieved through spacer layers of hexagonal boron nitride (hBN).
We record photoluminescence and reflection spectra at interlayer distances between 0.5~nm and 5.8~nm (0-16 hBN layers). 
We find that the energy transfer is dominated by states outside the light cone, indicative of a F\"orster transfer process, with an additional contribution from a Dexter process at 0.5~nm interlayer distance.   
We find that the measured dependence of the luminescence intensity on interlayer distances above 1~nm can be quantitatively described using recently reported values of the F\"orster transfer rates of thermalized charge carriers.
At smaller interlayer distances, the experimentally observed transfer rates exceed the predictions and furthermore depend on excess energy as well as on excitation density. 
Since the transfer probability of the F\"orster mechanism depends on the momentum of electron-hole pairs, we conclude that at these distances, the transfer is driven by non-relaxed charge carrier distributions.

\end{abstract}

\maketitle
\date{\today}

In low-dimensional systems, materials in proximity couple via near-field interactions, which can result in the transfer of energy from one material to the other \cite{Forster1949, Persson1982}. 
This interaction is not only central to light-driven processes in biological systems, including photosynthesis in plants \cite{Mirkovic2017}, but is also exploited in applications, such as organic light emitting diodes \cite{Yang2006} or in protein imaging \cite{Pollok1999}.
It furthermore appears in many artificial low-dimensional hybrid systems, such as quantum dots \cite{Chen2010}, semiconductor nano-platelets \cite{Federspiel2015} or molecular systems on graphene \cite{Treossi2009, Gaudreau2013}, as well as in van der Waals (vdW) heterostructures, e.g. of the transition metal dichalcogenides (TMDs) and graphene \cite{He2014_transfer,Hill2017,Froehlicher2018,Aeschlimann2020}.
These 2D heterostructures are of particular interest with respect to possible applications, such as flexible electronics \cite{Georgiou2013}, photodetectors \cite{Britnell2013, Zhang2014} or photovoltaics \cite{Bernardi2013,NassiriNazif2021}, the performance of which might be diminished or improved by energy transfer between the layers.
To take full advantage of the opportunities these heterostructures provide, a microscopic understanding of the mechanisms of the energy transfer is therefore of central importance \cite{Bradac2021}.

Different mechanisms can contribute to the transfer of energy.
The Dexter coupling describes electrons and holes which transfer independently due to an overlap of the respective wave functions. 
It therefore decreases exponentially with the distance between the materials \cite{Dexter1953, Bradac2021}. 
The F\"orster coupling is a dipole-dipole interaction, which is mediated by virtual photons and can therefore be significant also on larger interlayer distances \cite{Persson1982, Swathi2009}.
Theoretical works predict the F\"orster coupling to generally dominate in vdW heterostructures consisting of TMDs and  graphene \cite{Malic2014, Selig2019}.
Microscopically, these works predict the rates to display a strong dependence on in-plane momentum $Q$ of the excited electron-hole pairs, which vanishes for $Q=0$ \cite{Malic2014, Selig2019}. 
This particular momentum-dependence implies that the life time of excitons within the light cone does not depend on the total transfer rate $\gamma_t$ and, accordingly, the transfer to be dominated by excitations at larger momenta. 

Experimental studies in vdW heterostructures so far have focused on TMDs in direct contact with graphene.
Some have argued that the energy transfer in these heterostructures proceeds via tunneling \cite{Krause2021}, whereas others speculated the transfer to be dominated by a F\"orster interaction \cite{Froehlicher2018}.
Experiments conducted at large excitation densities were interpreted as being driven by a modified F\"orster transfer, facilitated by hot holes in graphene \cite{Dong2023}.
The dependence on interlayer distance has so far not been experimentally tested, and the techniques applied did not allow inferring information on the momentum-dependence of the transfer rates. 

The aim of this letter is to experimentally address the microscopic origins of the energy transfer in 2D heterostructures.
The distance dependence of the energy transfer rates is measured via the quenching of photoluminescence (PL) of \WS\ in the proximity to graphene using spacer layers of hBN of varying thickness \cite{Britnell2012, Brotons-Gisbert2019, Tebbe2023}. 
A dependence of the transfer rates on the momentum is observed in measurements of the line width of the luminescence, which originates from excitons within the light cone, as well as by varying the excitation condition in the PL experiments, which induces different transient electronic momentum distributions in the TMD and results in different levels of PL quenching.

%%%%%%%%%%%%%%%% Figure 1 %%%%%%%%%%%%%%%%%%%%%%%%%%%%%%%%%%%%%%%%%%%%%%%%%%%%%%%%%%%
\begin{figure}[!tbh]
    \begin{center}
    % reprint
        \includegraphics[width=1.0\columnwidth]{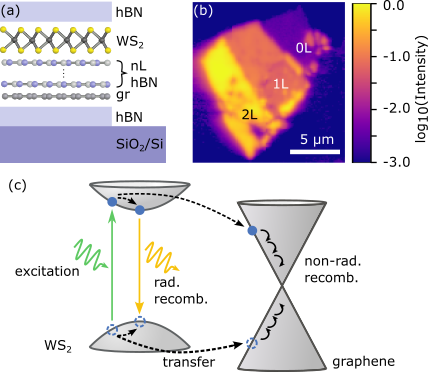}
        \caption{(a) Sketch of the \WS\ - graphene heterostructures used in this study. Both materials are separated by spacers of n layers (nL) of hBN. 
        (b) Photoluminescence image of a \WS\ - graphene heterostructue with 0, 1 and 2 hBN spacer layers. The photoluminescence intensity varies by approximately one and two orders of magnitude, respectively. 
        (c) Sketch of the competing electronic  processes: after excitation, electrons and holes can either recombine radiatively within \WS\, or transfer to graphene where they recombine non-radiatively. The relative efficiency of both processes determines the brightness of the emitted light from \WS. }
        \label{fig:imaging}
    \end{center}
\end{figure}
%%%%%%%%%%%%%%%%%%%%%%%%%%%%%%%%%%%%%%%%%%%%%%%%%%%%%%%%%%%%%%%%%%%%%%%%%%%%%%%%%%%%%

Heterostructures of \WS, hBN spacer layers and graphene were prepared by tape exfoliation of the individual materials, automated detection of suited flakes \cite{Uslu2023}, followed by mechanical stacking (see \cite{Tebbe2023} for details).  
In total, six different samples were studied in this work, with spacers ranging from zero to 16 layers of hBN.
This corresponds to distances between \WS\ and graphene of 0.5 nm to 5.8 nm, assuming a \WS-hBN  layer separation of 0.5 nm \cite{Rooney2017} and a graphene-hBN layer separation of 0.33 nm \cite{Pease1950, Haigh2012}.
A sketch of the sample structures is shown in \fig{fig:imaging} (a) and details on each sample are given in \cite{SI}.

All samples were characterized by taking high-resolution reflectance contrast as well as photoluminescence maps at room temperature using a hyperspectral imaging setup~\cite{Tebbe2023hyperspectral}.
An example of the spectrally integrated PL intensity of a sample containing 0, 1 and 2 hBN spacer layers (0L, 1L and 2L), excited with a 532 nm laser, is shown in \fig{fig:imaging} (b) (note the logarithmic intensity scale). 
The three areas can be distinguished in the raw image by the markedly different PL intensities, which differ by about one order of magnitude between adjacent spacer layer thicknesses. 
The magnitude of the PL quenching is a result of the competition between relaxation and radiative recombination of electron-hole pairs within \WS\ and energy transfer to graphene, which is followed by nonradiative recombination.
A sketch of the two competing processes is shown in \fig{fig:imaging} (c).

\begin{figure}[!tb]
    \begin{center}
    % reprint
        \includegraphics[width=1.0\columnwidth]{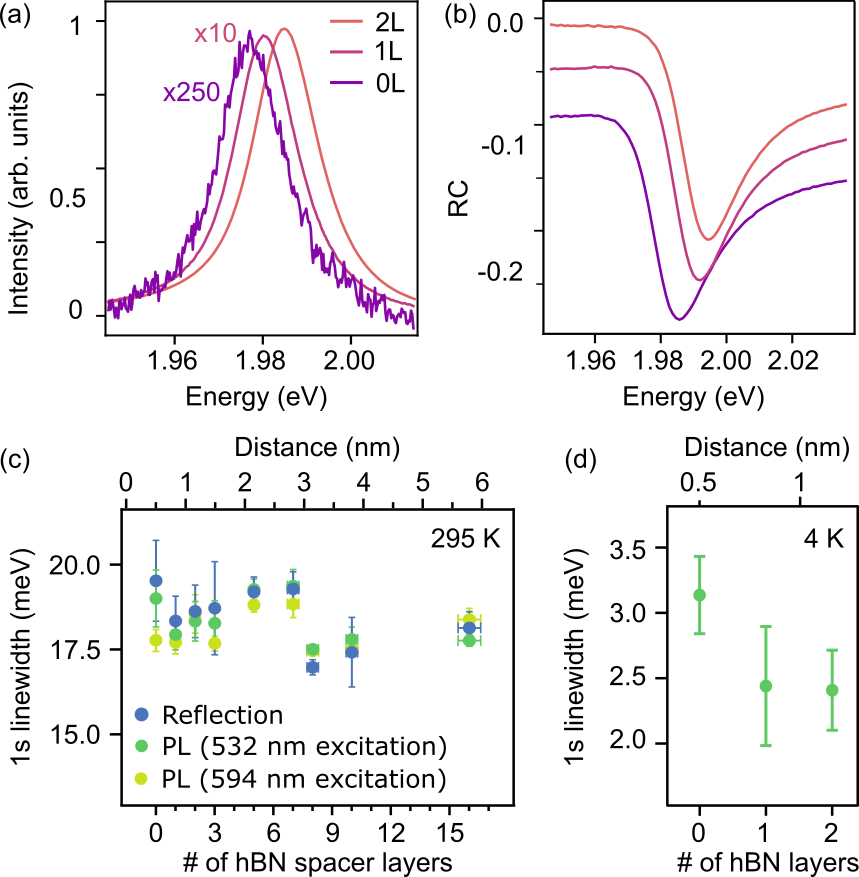}
        \caption{(a) PL spectra of three regions of a sample containing 0, 1 and 2 layers of hBN as spacers. 
        (b) Reflection contrast (RC) spectra from the same areas as in panel (a) (offset vertically for clarity).  
        (c) Minimal PL and RC line width at room temperature as a function of spacer layer thickness derived from all positions on all samples.  
        (d) Minimal PL Line width at small interlayer distances, obtained from one sample at 4K, at which phonon-related broadening is  reduced. 
        An increase of the line width of $(0.7 \pm 0.3)$ meV is observed for 0 spacer layers. }
        \label{fig:linewidth}
    \end{center}
\end{figure}

We first analyze the absorption and emission line widths in areas of different spacer layers, which are related to the lifetime of excitons within the light cone. 
Exemplary PL and reflection contrast spectra are shown in Figs. \ref{fig:linewidth} (a) and (b).
The PL spectra are well described by a single Lorentzian peak, except for areas without graphene, which show small signatures of additional trion emission \cite{Mak2013}.  
This suggests that the \WS\ crystal possesses some level of electron doping, and that these electrons transfer into the graphene \cite{Froehlicher2018} for all spacer layer thicknesses up to 5.8 nm.
The shape of the exciton absorption feature in the white light reflection contrast spectra is a result of thin film interference, which depends on the thicknesses of all layers of the sample and the substrate \cite{Arora2015}.
The redshift observed in both the absorption and emission peaks with proximity to graphene is due to the increasing dielectric screening from the graphene layer \cite{Raja2017, Tebbe2023}.

Fitting the emission and absorption spectra at all positions of all samples, we obtain distributions for the line width of every spacer layer.
From these, we extract the minimum line width and their error (see supplemental material for details \cite{SI}, which includes Refs. \cite{Tebbe2023, Li2014}). 
These values are shown as a function of the number of hBN spacer layers  in \fig{fig:linewidth}~(c). 

%%%%%%%%%%%%%%%% Figure 2 %%%%%%%%%%%%%%%%%%%%%%%%%%%%%%%%%%%%%%%%%%%%%%%%%%%%%%%%%%%
%%%%%%%%%%%%%%%%%%%%%ese%%%%%%%%%%%%%%%%%%%%%%%%%%%%%%%%%%%%%%%%%%%%%%%%%%%%%%%%%%%%%%%%

Within the errors of the experiment, absorption and emission line widths are found to be constant across all layer separations, with a possible increase only for the TMD in direct contact with graphene.
Since the minimum line width between samples can slightly depend on sample geometry, i.e. the top and bottom hBN thicknesses, due to the Purcell effect \cite{Fang2019}, the observed variations between layer thicknesses might originate from the different statistical weight of the various samples contributing to each data point.

To verify the broadening at small interlayer distances, we performed additional experiments on a single sample containing 0, 1 and 2 spacer layers, effectively eliminating variations due to the Purcell effect.
The experiments were done at a temperature of 4~K, at which phonon-related broadening is reduced, see \fig{fig:linewidth} (d).
These data yield a broadening of the exciton peak for 0 spacer layers of $(0.7\pm 0.3)$~meV compared to the peaks of 1 and 2 hBN layers, for which no significant difference in the FWHM is found.
We note that the broadening observed here is much smaller compared to earlier work, which found  an increase of $(5\pm2.5)$~meV when \WS\ was placed on graphene \cite{Hill2017}, but is similar to data on TMD/graphene heterostructures published more recently \cite{Lorchat2020}.
As in the former of the studies, samples were prepared on SiO$_2$, the broadening may not be systematic but likely resulted from inhomogeneities of the substrate \cite{Raja2019}. 

The broadening of the exciton peak allows first conclusions on the mechanism of the energy transfer. 
As the F\"orster transfer rates have been shown to approach zero for a momentum transfer of $Q=0$ \cite{Malic2014, Selig2019}, this interaction is not expected to affect excitons within the light-cone.
Our data therefore suggests that a Dexter process (simultaneous or consecutive transfer of electrons and holes) significantly contributes to the energy transfer at  0.5 nm interlayer distance (0 spacer layers), but not necessarily at larger interlayer distances.
As the  PL intensity is quenched by several orders of magnitude, while the observed broadening is smaller than the initial line width, we conclude that a different process is responsible for the majority of the energy transfer.

We now evaluate the quenching of the PL emission as a function of the graphene-\WS\ distance.
To experimentally cover the full range of distances, the PL intensities of different samples have to be compared to each other. 
Care needs to be taken in such a comparison, as absorption and emission of the sample are subject to thin film interference, which depends on the sample geometry, i.e. the order and thicknesses of the layers within the stack. 
We do account for this by comparing intensities of areas within the same samples, which differ by the number of spacer layers only. 
Intensities between samples are normalized by areas of common spacer thickness and sample and areas of line width exceeding 25 meV are excluded (see supplemental material \cite{SI}).

\begin{figure}[!tb]
    \begin{center}
    % reprint
        \includegraphics[width=1.0\columnwidth]{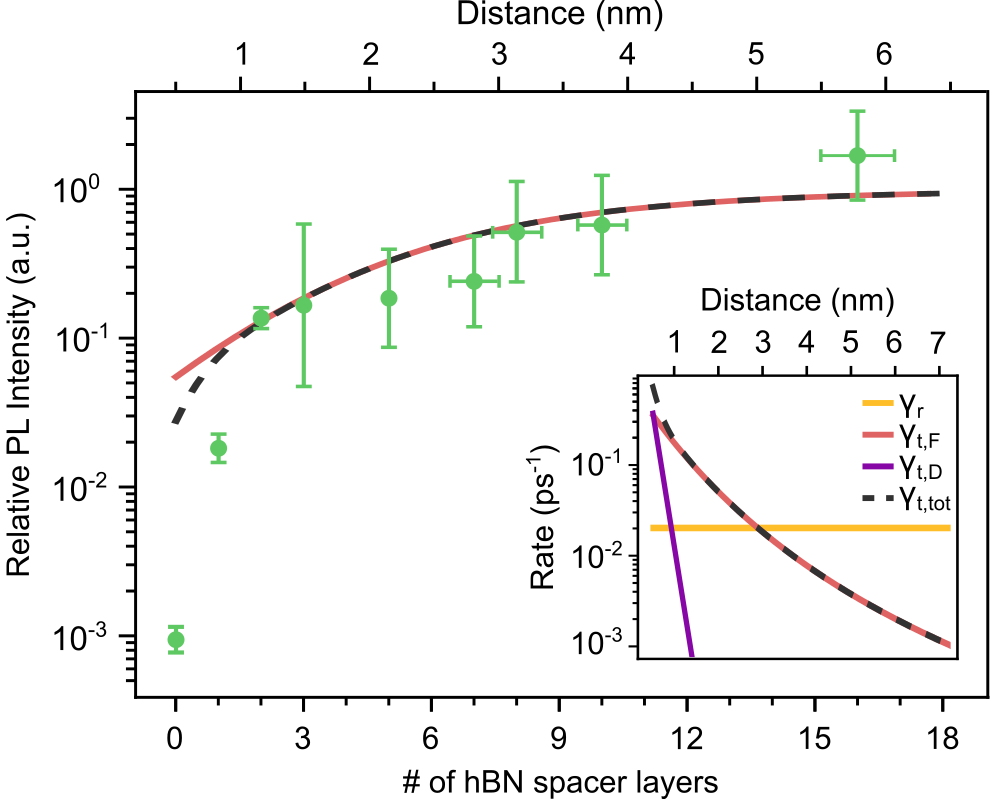}
        \caption{Dependence of the PL intensity on \WS-graphene interlayer distance (green data points). 
        The purple solid line is derived from the predicted F\"orster energy transfer rates at room temperature \cite{Selig2019} by varying the effective radiative rate to best reproduce the data above 1~nm interlayer distance. 
        The dashed black line additionally includes the experimentally obtained Dexter term. 
        The individual rates underlying the models are shown in the inset.}
        \label{fig:transfer}
    \end{center}
\end{figure}

The plot of the relative luminescence intensity vs spacer thickness is shown in Figure \ref{fig:transfer}. 
The measured PL intensity decreases with proximity to graphene, with the most drastic changes between layer thicknesses observed at interlayer distances smaller than 1~nm. 
In total, the intensity is reduced by more than three orders of magnitude between \WS\ in direct contact with graphene and the largest distance studied (5.8 nm). 
We also note that the PL intensity in areas without graphene is significantly smaller than the one at 16 spacer layers, which we attribute to the residual doping of these areas (see discussion above).

The measured luminescence intensity $I$ depends on the competing processes of radiative recombination and non-radiative transfer to graphene, with the associated rates $\gamma_r$ and $\gamma_t$, respectively.
Here we assume the radiative recombination rate to be independent of the interlayer distance, $\gamma_r(d)=\text{const}$.
Therefore, the PL intensity is connected to the rates by:
\begin{equation}
    I \propto ({\gamma_t(d)}/{\gamma_r} + 1)^{-1},
\end{equation}  
where $d$ denotes the interlayer distance \cite{Gomez-Santos2011}. 
Non-radiative decay within the TMD is also assumed to be unaffected by the presence of graphene; it then does not qualitatively influence the results and will be neglected in the following.

%%%%%%%%%%%%%%%% Figure 3 %%%%%%%%%%%%%%%%%%%%%%%%%%%%%%%%%%%%%%%%%%%%%%%%%%%%%%%%%%%
%%%%%%%%%%%%%%%%%%%%%%%%%%%%%%%%%%%%%%%%%%%%%%%%%%%%%%%%%%%%%%%%%%%%%%%%%%%%%%%%%%%%%

The connection between transfer rates and luminescence intensity allows us to compare our measurements to a recent theory of the distance dependence of the F\"orster energy transfer between \WS\ and graphene \cite{Selig2019}.
The transfer probabilities of charge carriers have been shown to depend on interlayer distance and the total momentum $Q$ of the electron-hole pairs as 
\begin{equation}
    \gamma_{t,F}(Q,d) \propto Q^2 \cdot e^{-2 Qd}.
\end{equation}
Assuming the carriers to be thermalized, i.e. them following a Boltzmann distribution \cite{Brem2018}, an effective transfer rate can then be calculated. 
Since the increased line width at 0 spacer layers demonstrates a contribution from Dexter coupling, which is neglected in the theory, we also add a phenomenological Dexter term, such that $\gamma_t(d)=\gamma_{t,F}(d)+\gamma_{t,D}(d)$. 

We calculate the F\"orster rates using the parameters given in \cite{Selig2019} with an effective (i.e. frequency-averaged) dielectric constant of hBN of $\epsilon_{\textrm{eff}}=4.5$  \cite{Steinhoff2018, Waldecker2019}, in which our samples are encapsulated. 
The Dexter transfer rate at interlayer distances of 0.5 nm is extracted from the linewidth broadening of $(0.7 \pm 0.3)$ meV, corresponding to $\gamma_{t,D}(0.5 \mathrm{nm}) = 0.18 \pm 0.08$ ps$^{-1}$.
We further assume that it decreases exponentially with distance by one order of magnitude per hBN layer \cite{Britnell2012,Bradac2021}.
To compute the expected PL quenching, the radiative rate is adjusted as the only free parameter.
We find that for $\gamma_r = 0.02$ ps$^{-1}$, which is close to values reported for \WS\ encapsulated in hBN at room temperature \cite{Fu2019}, the F\"orster transfer alone reproduces the data for interlayer distances larger than 1~nm, see the solid line in \fig{fig:transfer} (the Dexter transfer is negligible at these distances). 
For 0 and 1 spacer layers, however, the F\"orster mechanism fails to describe the quenching by a large margin.
Including the Dexter transfer (dashed line) does increase the predicted PL quenching, but can not account for the large discrepancy.

\begin{figure}[!tb]
    \begin{center}
    % reprint
        \includegraphics[width=1.0\columnwidth]{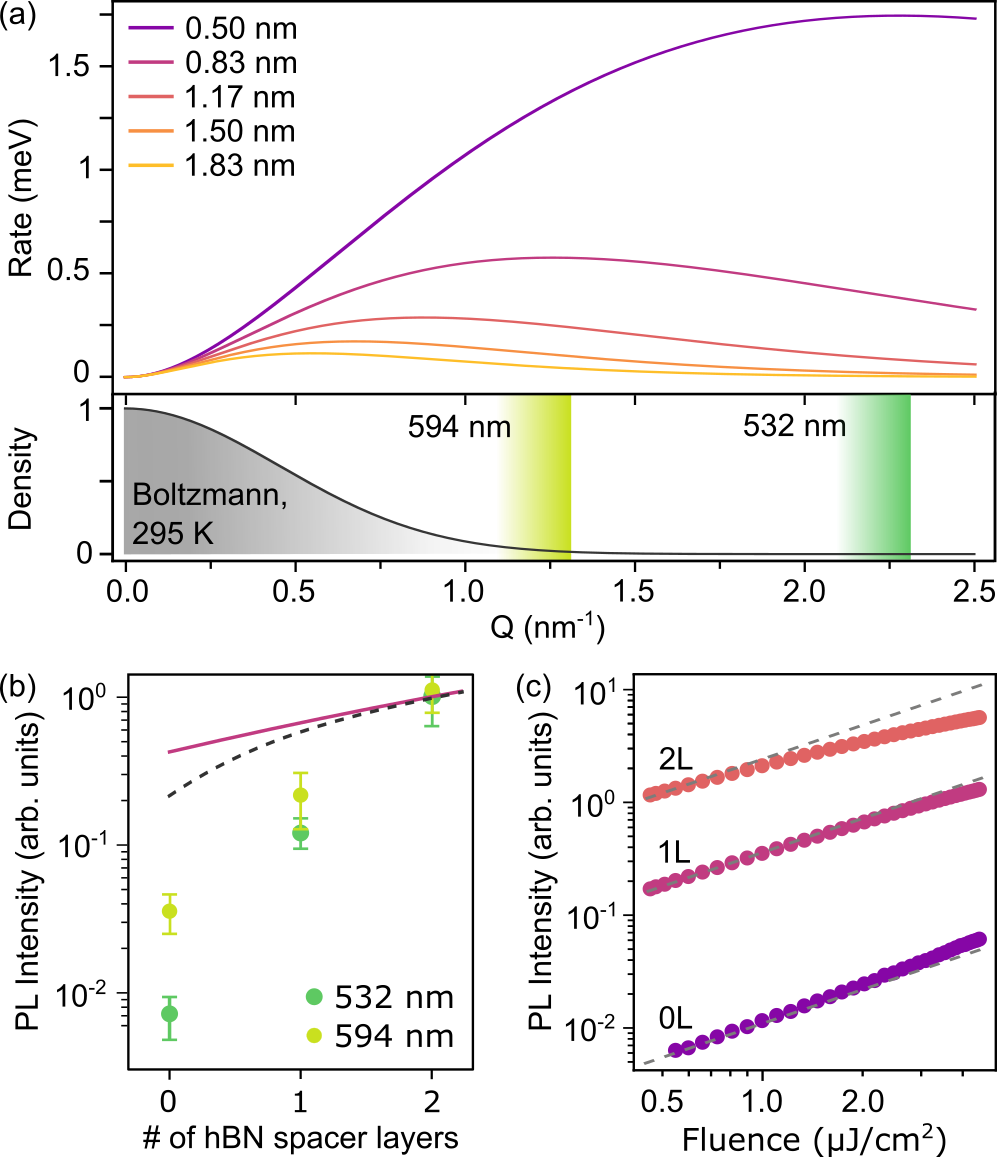}
        \caption{(a) Calculated broadening due to F\"orster energy transfer as a function of momentum transfer $Q$ for different interlayer distances. Bottom: Boltzmann distribution at 295 K and maximum available momenta for cw excitation with a 532 nm laser (green) and a 594 nm laser (yellow). 
        (b) Photoluminescence intensity in areas of 0, 1 and 2 spacer layers for an excitation wavelength of 532nm (green) and 594 nm (yellow). 
        (c) Power dependence of the PL intensity for 0, 1 and 2 spacer layers. 
        The dashed gray lines indicate a linear increase. }
        \label{fig:nonthermal}
    \end{center}
\end{figure}

We next discuss possible reasons for the increased PL quenching at small interlayer distances. 
In \fig{fig:nonthermal} (a), the momentum-dependence of the F\"orster energy transfer rates is shown at various interlayer distances.
At the smallest distances, the transfer rates at large momenta become comparable to typical relaxation rates (thermalization and cooling), on the order of one to few tens of picoseconds \cite{Robert2016, Brem2018, Wang2023}. % after few 
As electron-hole pairs created with sufficient excess energy can acquire a finite center-of-mass momentum due to scattering with other carriers or with phonons before they are relaxed \cite{Brem2018, Selig2019dynamics, Waldecker2017}, we speculate that the measured transfer rates originate from non-relaxed carrier distributions.

This supposition is corroborated by investigations of the PL quenching using different excitation conditions.
For the 532 nm (2.33 eV) laser, electron-hole pairs are created with an excess energy of approximately 340 meV compared to the exciton peak at 1.99 eV. 
Using an exciton effective mass of 0.6 $m_e$ \cite{Selig2019}, the 532 nm  laser allows excitations to acquire momenta of up to $Q\approx2.3$ nm$^{-1}$.
A second cw laser with a wavelength of 594 nm (2.09 eV) creates carriers with a maximum accessible momentum of $Q\approx1.3$ nm$^{-1}$.
These momenta are indicated in \fig{fig:nonthermal}~(a) by the green and yellow gradients.

%%%%%%%%%%%%%%%% Figure 4 %%%%%%%%%%%%%%%%%%%%%%%%%%%%%%%%%%%%%%%%%%%%%%%%%%%%%%%%%%%
%%%%%%%%%%%%%%%%%%%%%%%%%%%%%%%%%%%%%%%%%%%%%%%%%%%%%%%%%%%%%%%%%%%%%%%%%%%%%%%%%%%%%

\fig{fig:nonthermal} (b) shows a comparison of the PL intensities for interlayer distances $\le 1$~nm for excitation with the two laser sources at the same incident power.
An increase of PL intensity of a factor of five is observed for the excitation with the smaller excess energy.
This corresponds to a significant reduction of the total transfer rate, which demonstrates that, indeed, the charge carrier distributions substantially influence the effective transfer rates. 

Lastly, we describe the quenching at higher excitation densities, at which interactions are expected to play a role, and which are typically employed in time-resolved experiments. 
We use a supercontinuum laser, filtered to a wavelength of (532$\pm5)$ nm, which has a pulse duration of approximately 10~ps and a repetition rate of 80 MHz.
In \fig{fig:nonthermal} (c), the PL intensities at 0, 1 and 2 spacer layers are shown as a function of excitation fluence. 
For two and one spacer layers, we observe the onset of a sub-linear dependence, most likely caused by  exciton-exciton annihilation \cite{Yuan2015}.
In the region of direct contact (0L), however, the intensity increases slightly super-linearly. 
We speculate that this is caused by an increased scattering rate, which allows some carriers to reach the light cone that would otherwise have been transferred to graphene. 
These data not only corroborate that carrier dynamics within \WS\ play an important role in the energy transfer process, but also show that transfer rates, measured in time-resolved experiments, might deviate significantly from those in the continuous excitation regime.

The picture which emerges from our study is that, at room temperature, and for interlayer distances above approximately 1~nm,  energy transfer is dominated by F\"orster interactions.
The effective transfer rates in this regime have been calculated by thermally averaging the individual (momentum-dependent) transfer probabilities and well reproduce the data.
Therefore, in this regime, the well known distance$^{-4}$ relation of the transfer rates should hold.
Below 1~nm interlayer distance, the individual F\"orster transfer rates become large enough that on average, charge carriers transfer before they fully relax.
The total transfer rate is then highly dependent on the excitation condition and is affected by the excess energy as well as the excitation density. 
A contribution from a Dexter transfer is observed at the smallest interlayer distances, and we infer its magnitude to become comparable to the F\"orster rates only for the two materials in direct contact.
We note that the Dexter contribution might depend on the twist angle between the layers.
Since our samples were not purposefully aligned, however, we did not observe significant differences between samples. 

While we are able to identify and demonstrate different mechanisms leading to the energy transfer at different interlayer distances, below 1~nm, the experimentally observed PL quenching is still larger than expected for the calculated transfer probabilities in \cite{Selig2019}, even assuming unrealistic carrier distributions.   
One possibility is that dark excitonic states affect the total transfer rates \cite{Lin2023}, which were not explicitly included in the theory. 
It is also possible that a Meitner-Auger type mechanisms contributes to the total transfer, even though the highest excitation densities here are approximately 5 orders of magnitude lower than in the work introducing this mechanism \cite{Dong2023}.
Further theoretical and experimental studies, which explicitly take into account the dynamics of non-relaxed carriers, are needed to clarify these details. 

To conclude, our results establish a picture of the dominant mechanisms of energy transfer in vdW heterostructures at different interlayer distances, which can serve as a basis for the development of devices requiring the transfer of charge and energy, such as optical detectors or solar panels.

The supporting data for this article are openly available from zenodo \cite{zenodo}.

%\section{Acknowledgements}
The authors thank Ermin Malic and Joshua Thompson for fruitful discussions. 
This project has received funding from the European Union’s Horizon 2020 research and innovation programme under grant agreement No 881603, by the Deutsche Forschungsgemeinschaft (DFG, German Research Foundation) under Germany's Excellence Strategy - Cluster of Excellence Matter and Light for Quantum Computing (ML4Q) EXC 2004/1 – 390534769. K.W. and T.T. acknowledge support from the JSPS KAKENHI (Grant Numbers 20H00354, 21H05233 and 23H02052) and World Premier International Research Center Initiative (WPI), MEXT, Japan.

\end{document}

\title{Supplementary Material: Distance dependence of the energy transfer mechanism in \WS\ - graphene heterostructures}
\author{David Tebbe} 
\email{david.tebbe@rwth-aachen.de}
\author{Marc Sch\"utte} 
%\author{Alexander Polkowski} 
\affiliation{$2^{nd}$ Institute of Physics, RWTH Aachen University, 52074 Aachen, Germany}
\author{K.~Watanabe}
\affiliation{Research Center for Functional Materials, National Institute for Materials Science, 1-1 Namiki, Tsukuba 305-0044, Japan}
\author{T.~Taniguchi}
\affiliation{International Center for Materials Nanoarchitectonics, National Institute for Materials Science, 1-1 Namiki, Tsukuba 305-0044, Japan}
\author{Christoph Stampfer} 
\affiliation{$2^{nd}$ Institute of Physics, RWTH Aachen University, 52074 Aachen, Germany}
\affiliation{Peter Gr\"unberg Institute (PGI-9), Forschungszentrum J\"ulich, 52425 J\"ulich, Germany}
\author{Bernd Beschoten} 
\affiliation{$2^{nd}$ Institute of Physics, RWTH Aachen University, 52074 Aachen, Germany}
\affiliation{Peter Gr\"unberg Institute (PGI-9), Forschungszentrum J\"ulich, 52425 J\"ulich, Germany}
\author{Lutz Waldecker} 
\email{waldecker@physik.rwth-aachen.de}
\affiliation{$2^{nd}$ Institute of Physics, RWTH Aachen University, 52074 Aachen, Germany}

\maketitle

\section{Sample geometries}

This study was performed on a total of six samples with different combinations of spacer layer thicknesses. 
While similar top- and bottom hBN flakes were chosen, the sample geometries vary between samples; an overview is given in Supplementary Table \ref{tab:samples}.
The top and bottom hBN thicknesses were determined by atomic force microscopy and the number of spacer layers was determined from their optical contrast and atomic force microscopy. 
All samples were placed on Si/SiO$_2$ substrates with an oxide thickness of 285 nm.

\begin{table}[h]
\begin{ruledtabular}
    \begin{tabular}{ccccc}
       
     sample & top hBN  & bottom hBN  & backgate  & \# hBN spacers  \\
            & thickness (nm) & thickness (nm) & thickness (nm) &   \\
       \hline
     S1  & 41 & 35  & 0 &   0,1,2 \\ % 0906a
     S2  & 38  & 42  & 0 &   0,2,ng \\ % 0510a
     S3  & 25  & 36  & 2.7 &   2,3 \\ % Optimus
     S4  & 20  & 37  & 0 &   2,5,7,8,10 \\ % 0601c  
     S5  & 28  & 34  & 2 &   1,2 \\ % Survivor
     S6  & 40   & 45   & 0 &   0,16 \\ %0915b
    % S7 0915c & 28 (31) & 35 (54) &   &   0,2,ng \\
    % S8 0428b  &   &   &   &   0,2,ng \\
    \end{tabular}
    \caption{Overview of the geometry of all samples used in this study.  
    \label{tab:samples}
        }
\end{ruledtabular}
\end{table}

PL images of all samples, obtained by integrating the PL spectra from 1.9 eV to 2.05 eV, are shown in Supplementary Fig. \ref{fig:SI_samples}.
Note that the spatial resolution in x and y direction of the raw images was not the same due to experimental constraints and therefore appear stretched; maps shown in the main test were corrected for this distortion.

%%%%%%%%%%%%%%%% Figure SI1 %%%%%%%%%%%%%%%%%%%%%%%%%%%%%%%%%%%%%%%%%%%%%%%%%%%%%%%%%%%
\begin{figure}[!htb]
    \begin{center}
    % reprint
        \includegraphics[width=0.85\columnwidth]{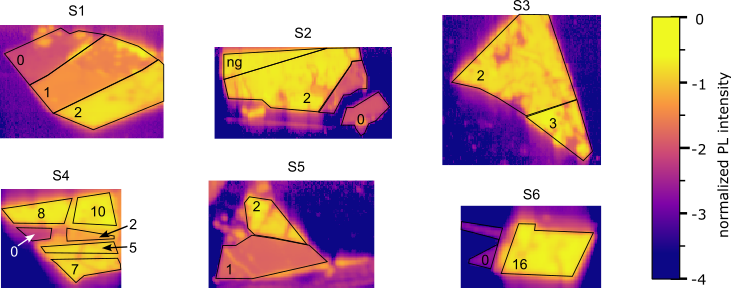}
        \caption{a) PL images of all samples used in this study. For better visibility of the 0 layer areas, images excited with the 594 nm laser are shown.  }
        \label{fig:SI_samples}
    \end{center}
\end{figure}
%%%%%%%%%%%%%%%%%%%%%%%%%%%%%%%%%%%%%%%%%%%%%%%%%%%%%%%%%%%%%%%%%%%%%%%%%%%%%%%%%%%%%

\section{Analysis of photoluminescence intensity}

\begin{figure}[!htb]
    \begin{center}
    % reprint
        \includegraphics[width=\columnwidth]{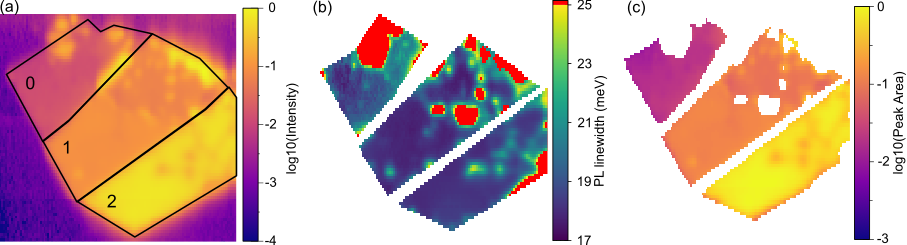}
        \caption{a) PL image and b) fitted line width of sample S1. c) Fitted Peak area of sample S1. Pixels at which the line width exceeds 25 meV are excluded from the analysis.}
        \label{fig:SI_PL}
    \end{center}
\end{figure}

All photoluminescence (PL) intensities presented in the main text are averages extracted from PL maps, similar to those shown in Supplementary Fig. \ref{fig:SI_samples}.
In a first step, PL intensities of each individual sample area are averaged.
In these averages, we exclude pixels at which the line width exceeds 25 meV (see Fig. Supplementary Figure \ref{fig:SI_PL}), which mostly correspond to positions with visible bubbles in microscope images and therefore to sample areas with obvious inhomogeneities.  

To average the PL intensities of different samples, we normalize the intensities of the individual samples to areas of common spacer thickness.  
This is necessary as the intensity of the excitation laser on the sample, as well as the emission of the luminescence light towards the microscope objective, depend on the sample geometry (in particular on top and bottom hBN thicknesses) due to thin film interference.
Since five out of six sample contain an area of 2 spacer layers, these are used for normalization.
The remaining sample is compared to all others by matching the intensity of the area without spacer layer (0 layers).

%%%%%%%%%%%%%%%% Figure SI2 %%%%%%%%%%%%%%%%%%%%%%%%%%%%%%%%%%%%%%%%%%%%%%%%%%%%%%%%%%%
%%%%%%%%%%%%%%%%%%%%%%%%%%%%%%%%%%%%%%%%%%%%%%%%%%%%%%%%%%%%%%%%%%%%%%%%%%%%%%%%%%%%%

\section{Analysis of the minimal line width}

All emission spectra were fitted by a Lorentzian peak, from which the line width as well as the position and intensity are obtained.
The reflection contrast data were fitted by constructing a dielectric function for the \WS\ layer from Lorentz oscillators \cite{Li2014} and calculating the respective reflection spectrum with the transfer matrix method (details can be found in \cite{Tebbe2023}).  

From the fits of the RC and PL spectra, we obtain distributions of the line width, which have a steep rise on the low-energy side and a longer tail on the high energy side (see Supplementary Figure \ref{fig:SI_histogram}).
Assuming that samples have a homogeneous linewidth, which is only broadened but never narrowed by external inhomogeneities, such a distribution is expected. 
We therefore approximate these distributions with an exponentially modified Gaussian function:
\begin{equation}
    f(x,\mu,\sigma,\lambda)=\frac{\lambda}2 e^{\frac{\lambda}2(2\mu + \lambda\sigma^2-2x)}  \mathrm{erfc} \left(\frac{\mu+\lambda\sigma^2-x}{\sqrt{2}\sigma} \right).
\end{equation}
In this case, the exponentially decaying part corresponds to the distribution of the line width of the sample and the Gaussian part to the uncertainties introduced by the measurement. 
The homogeneous line width is therefore extracted as the parameter $\mu$ and the error of the minimal line width is given by $\sigma$, i.e. the width of the Gaussian part.

%%%%%%%%%%%%%%%% Figure SI2 %%%%%%%%%%%%%%%%%%%%%%%%%%%%%%%%%%%%%%%%%%%%%%%%%%%%%%%%%%%
\begin{figure}[!htb]
    \begin{center}
    % reprint
        \includegraphics[width=0.5\columnwidth]{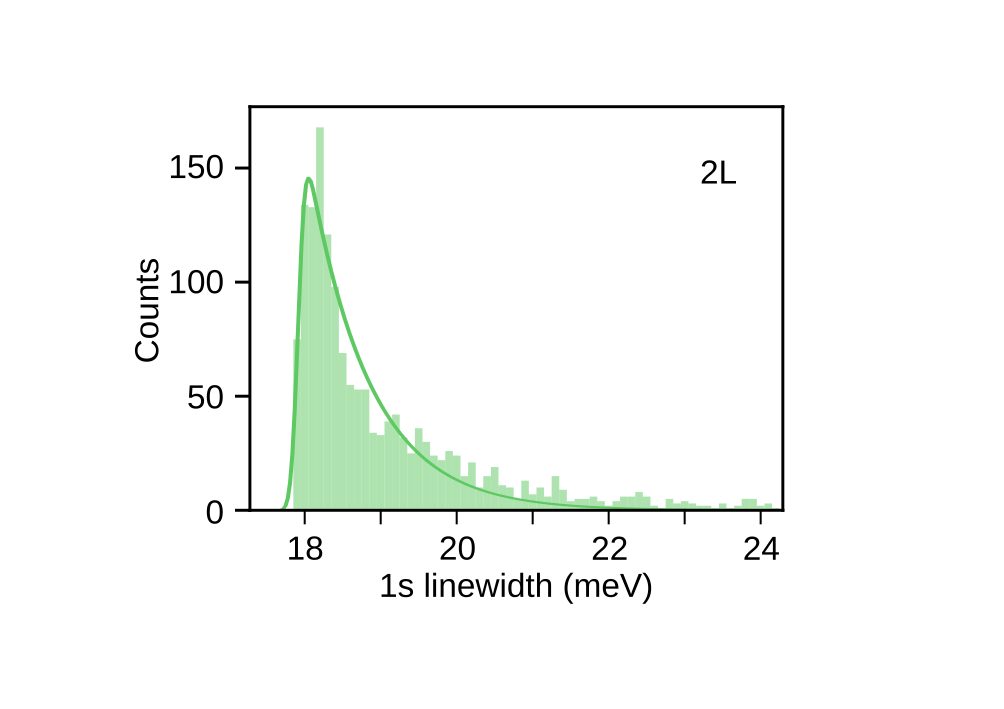}
        \caption{Histogram of the fitted line width of a sample area containing two hBN spacer layers.  }
        \label{fig:SI_histogram}
    \end{center}
\end{figure}
%%%%%%%%%%%%%%%%%%%%%%%%%%%%%%%%%%%%%%%%%%%%%%%%%%%%%%%%%%%%%%%%%%%%%%%%%%%%%%%%%%%%%

\section{Rate equation model for PL quenching}

Without loss of generality, we assume an instantaneous excitation of a number of carriers $N_0$ at time $t=0$.
As these start to decay, the number of excited carriers at time $t$  will be $N(t)$. 
The measured photoluminescence intensity in this scenario is proportional to the number of recombined carriers at very long delays $N_r(t=\infty) \equiv N_r$. 
Note that the total number of radiatively recombined carriers $N_r$ and transferred carriers $N_t$ add up to $N_r + N_t = N_0$. 
Let $\gamma_i=1/\tau_i$ ($i=r,t$) be the respective  rates and life times of radiative recombination and transfer.
We can then set up a set of rate equations to describe the evolution of each population as:
\begin{align}
\frac{dN}{dt} &= -N(\frac1{\tau_r}+\frac1{\tau_t}), \\
\frac{dN_r}{dt} &= \frac{N}{\tau_r}, \\
\frac{dN_t}{dt} &= \frac{N}{\tau_t}.
\end{align}

These rate equations have solutions
\begin{align}
N(t) &= N_0 \cdot e^{-t(\frac1{\tau_r}+\frac1{\tau_t})} \\
N_r(t) &= N_0 \cdot \frac1{1+\frac{\tau_r}{\tau_t}} \cdot (1-e^{-t(\frac1{\tau_r}+\frac1{\tau_t})}) \\
\end{align}

At $t=\infty$, we then obtain
\begin{align}
\frac{N_r}{N_0} = \frac1{1+\frac{\tau_r}{\tau_t}} = \frac1{1+\frac{\gamma_t}{\gamma_r}}.
\end{align}

Similarly, $\frac{N_t}{N_0} =  \frac1{1+\frac{\gamma_r}{\gamma_t}}$.
\\

\newpage
\section*{References}